\documentclass{ametsocV6.1}

\nolinenumbers

\title{The Link between Gulf Stream Precipitation and European Blocking in General Circulation Models and the Role of Horizontal Resolution}

\authors{Kristian Strommen\aff{a}\correspondingauthor{Kristian Strommen, kristian.strommen@physics.ox.ac.uk},
Simon L. L. Michel\aff{a},
Hannah M. Christensen\aff{a}
}

\affiliation{\aff{a}{Department of Physics, University of Oxford}}

\abstract{Past studies show that coupled model biases in European blocking and North Atlantic eddy-driven jet variability decrease as one increases the horizontal resolution in the atmospheric and oceanic model components. 
This has commonly been argued to be related to an alleviation of sea surface temperature (SST) biases due to increased oceanic resolution in particular, with a physical pathway via changes to surface baroclinicity.
On the other hand, many studies have now highlighted the key role of diabatic processes in the Gulf Stream region on blocking formation and maintenance. Here, following recent work by Schemm, we leverage a large multi-model ensemble to show that Gulf Stream precipitation variability in coupled models is tightly linked to the simulated frequency of European blocking and northern jet excursions. Furthermore, the reduced biases in blocking and jet variability are consistent with greater precipitation variability as a result of increased atmospheric horizontal resolution. By contrast, typical North Atlantic SST biases are found to share only a weak or negligible relationship with blocking and jet biases. Finally, while previous studies have used a comparison between coupled models and models run with prescribed SSTs to argue for the role of ocean resolution, we emphasise here that models run with prescribed SSTs experience greatly reduced precipitation variability due to their excessive thermal damping, making it unclear if such a comparison is meaningful. Instead, we speculate that most of the reduction in coupled model biases may actually be due to increased atmospheric resolution.}

\begin{document}

\maketitle

%
%
%
%
%
%

%








\section{Introduction}

Accurately simulating the variability of the North Atlantic eddy-driven jet and European blocking events has been a longstanding challenge in climate science and meteorology. Considerable progress has been made on this challenge in the last 30 years \citep{davini2016northern, davini2020cmip3, DorringtonStrommenFabiano2022}, at least partially as a result of the increased horizontal resolution of models \citep{davini2020cmip3, schiemann2020northern}. However, models still exhibit systematic biases. A notable and well known example is that models still underestimate the occurrence of (a) European blocking events \citep{davini2020cmip3, schiemann2020northern}, and, relatedly, (b) northern jet excursions \citep{DorringtonStrommenFabiano2022}, leading to uncertainty in future projections of how such events will change with global warming \citep{woollings2018blocking}. Weather forecasts also still struggle to accurately predict European blocking events \citep{matsueda2018estimates, bueler2021year}. Since European blocking events are closely linked to extreme Euro-Atlantic surface weather \citep{kautz2022atmospheric}, a lot of effort has been devoted to understanding the origin of model biases and the exact role played by model resolution. The purpose of this paper is to contribute towards this effort. For pointers towards more theoretical work on blocking, the reader may refer to, e.g., \citet{woollings2018blocking}.

Much of the past literature on biases in European blocking and North Atlantic jet variability has focused on the role of North Atlantic sea surface temperature (SST) biases in models, especially in the Gulf Stream region. The separation of the Gulf Stream from the North American continent creates a sharp SST gradient that can affect downstream jet and blocking variability by modulating surface wind convergence \citep{minobe2008influence}, surface baroclinicity \citep{brayshaw2011basic} and meridional heat transport \citep{ambaum2014nonlinear, o2016influence, novak2017marginal, o2017gulf}. Following its separation from the continent, the Gulf Stream current turns northwards due to bottom topography, before veering east again towards Europe (the so-called ``northwest corner''), with important consequences for the mean state of the SSTs in the central North Atlantic \citep{Keeley2012, drews2015use}. However, the Gulf Stream in models fails to separate early enough from the continent \citep{chassignet2008gulf, moreno2022impact, tsartsali2022impact}, leading to an overly smooth SST gradient \citep{o2016influence}. Models also fail to simulate the ``northwest corner'', and instead simulate a Gulf Stream current that extends zonally from Newfoundland to Europe, resulting in North Atlantic SSTs exhibiting a cold bias that can be as great as $10 ^{\circ}C$ \citep{Keeley2012, wang2014global, drews2015use}. Recently, \citet{athanasiadis2022mitigating} also argued that models show an overly \emph{sharp} SST gradient in the central North Atlantic compared to observations\footnote{While \citet{athanasiadis2022mitigating} do not comment on the explicit source of this bias, it appears to be a natural consequence of the failure of models to simulate the ``northwest corner''. We return to this in the Discussion.}, and that the associated bias in surface baroclinicity is partly responsible for European blocking biases.

Because it is known that correctly simulating the pathway of the Gulf Stream depends at least in part on the ocean model resolution \citep{chassignet2008gulf, tsartsali2022impact}, it is natural to hypothesise that much of the bias in European blocking and jet variability is a result of insufficient horizontal resolution in models, especially in the ocean component. This hypothesis has been explicitly made and argued for in some select single-model studies, such as \citet{scaife2011improved}, and more recently in the three multimodel studies \citet{davini2020cmip3}, \citet{athanasiadis2022mitigating} and \citet{michel2023increased}. Since the atmospheric and oceanic horizontal resolutions are typically increased in tandem, a clean attribution of their relatives roles cannot be immediately made. Nevertheless, \citet{scaife2011improved}, \citet{athanasiadis2022mitigating} and \citet{michel2023increased} all argue that ocean resolution is more important, with a key piece of evidence in all cases being a comparison between coupled models and models forced with prescribed observed SSTs (which we henceforth refer to as AMIP models). Specifically, they find that the bias reduction seen with increased resolution is either absent or much smaller in AMIP models compared to coupled models, and interpret this as evidence for the importance of ocean resolution. However, the evidence here does not seem completely conclusive. Firstly, the classic work of \citet{barsugli1998basic} on air-sea coupling suggests that differences between AMIP and coupled models could be hard to interpret, due to the excessive thermal damping experienced by AMIP models. 
More recently, \citet{mathews2024oceanic} have emphasised the role of air-sea coupling to blocking maintenance, which, if correct, would result in further systematic differences between AMIP and coupled models.
Secondly, AMIP models broadly have similar levels of biases in blocking compared to coupled models \citep{schiemann2020northern}, despite having no SST biases whatsoever. Thirdly, some of the earlier studies arguing for the role of the ocean are based on small ensembles, such as \citet{scaife2011improved} (1 ensemble member), which we now know to be inadequate in the face of the small signals and large internal variability characteristic of the North Atlantic circulation \citep{scaife2018signal}. In some cases, it is known that hundreds of ensemble members are needed to robustly detect forced signals in the North Atlantic \citep{ye2024response}, making it unclear if even the 17 ensemble members used in \citet{athanasiadis2022mitigating} are sufficient.

On the other hand, a growing body of literature has highlighted the crucial role played by \emph{diabatic processes} in the Gulf Stream region in the formation and maintenance of blocking \citep{pfahl2015importance, steinfeld2019role, steinfeld2020sensitivity, yamamoto2021oceanic, wenta2024linking, mathews2024oceanic} and the North Atlantic storm track \citep{marcheggiani2023diabatic}. In this vein, \citet{schemm2023toward} performed AMIP simulations with a regionally varying horizontal atmospheric resolution, where the resolution is around 10km in the storm track region and around 5km in the Gulf Stream region. Schemm finds that this increased resolution leads to enhanced precipitation rates in the Gulf Stream region, greater northward propagation of cyclones, and an associated jet which is more tilted and northwards. Schemm therefore hypothesises that the ``too zonal, too equatorwards'' bias in the North Atlantic jet seen in models is a direct result of inadequately resolved diabatic processes in the Gulf Stream region, and that this can be alleviated with increased horizontal resolution. While Schemm does not explicitly discuss European blocking, it is natural, based on the crucial role played by cyclones in creating and maintaining blocking anticyclones \citep{colucci1985explosive, nakamura1993synoptic, yamazaki2009selective, wenta2024linking}, to hypothesise that this could also explain model biases in European blocking. The recent work of \citet{mathews2024gulfstream} adds compelling evidence towards this, by showing that forcibly suppressing the Gulf Stream moisture fluxes in a model leads to a large reduction in blocking frequencies across the entire northern hemisphere.

In this study, we aim to test Schemm's hypothesis against a large coupled multimodel ensemble (135 members), spanning a wide range of model resolutions (from 400 to 25km atmospheric grid spacing at the equator). Concretely, we will use the ensemble to address the following questions:

\begin{itemize}
    \item[1.] Is Gulf Stream precipitation variability clearly linked to the mean frequency of occurrence of European blocking events and northward eddy-driven jet excursions in models? 
    \item[2.] To what extent does atmospheric horizontal resolution affect Gulf Stream precipitation variability, and can the improved representation of European blocking and northward jet excursions with model resolution be explained by improved precipitation variability?
\end{itemize}

\noindent We emphasise immediately that measuring precipitation \emph{variability}, as opposed to just mean precipitation, is an important part of our approach. This can be motivated in several ways, including the observation that the heatflux variability in storm tracks happens in intermittent `bursts' \citep{messori2013sporadic, ambaum2014nonlinear}, an important physical feature which can only be captured using moments beyond the mean. Further motivation is given in the Methods section.

Additionally, drawing on \citet{barsugli1998basic}, we will examine a third point:
\begin{itemize}
    \item[3.] How does Gulf Stream precipitation variability compare between coupled and AMIP models, and what can we say about the relative role of atmospheric vs oceanic resolution in reducing biases in European blocking?
\end{itemize}
As part of our discussion on the role of atmospheric versus oceanic resolution, we also look at how SST biases in our ensemble are linked to blocking, including revisiting the hypothesis of \citet{athanasiadis2022mitigating}.

We emphasise up front that the purpose of this paper is not to perform in-depth analysis of how physical processes are or are not represented across the 135 model simulations. We leave this as an important avenue for future work.

\section{Data}
\label{sec:data}

We analyse data drawn from a total of 135 historical model simulations, consisting of multiple ensemble members drawn from a range of CMIP6 models \citep{Eyring2016} CMIP5 models \citep{Taylor2012}, and HighResMIP models \citep{Haarsma2016}. The CMIP5 and CMIP6 historical experiments consist of coupled uninitialised climate runs forced with historical greenhouse gas and aerosol forcings over the 20th century, after a spin-up from a free-running pre-industrial control run. The CMIP5 simulations span either 1900-2005 or 1950-2005, while the CMIP6 simulations span 1900-2015 or 1950-2015. The HighResMIP models are initialised in 1950, following a short 50-year spin-up and span the 65 years between 1950 and 2015, and use the same historical forcings as CMIP6. The data is in all cases restricted to the December-January-February seasons between 1979-01 and 2014-12. Thus CMIP5 data covers 1979-2005 and CMIP6 and HighResMIP cover 1979-2014. The HighResMIP models are sometimes referred to as the ``PRIMAVERA'' models, in reference to the European Union Project under which they were produced. Tables detailing the exact models and ensemble members used are given in the Supplementary Information (Tables S1, S2 and S3). 

As our observational estimate, we use ERA5 \citep{Hersbach2020} for SSTs, 850 hPa winds (U850) and 500 hPa geopotential height (Z500). For precipitation, it is crucial to avoid using precipitation from ERA5. Reanalysis rainfall is based on model output, and thus suffers from the typical "drizzle bias" of models, whereby they overestimate low precipitation events and underestimate high precipitation events \citep{chen2021convective}. Instead, we use the NASA/Goddard Space Flight Center satellite product IMERG version 7, hereafter IMERG \citep{huffman2020integrated, huffman2023imerg}. This is a greatly updated version of the now defunct TRMM/3B42 satellite data set \citep{kawanishi2000trmm}. IMERG has a native spatial resolution of 0.1° (10km) and estimates the rainfall every 30 minutes by post-processing the output of a number of satellites. Extensive discussion on the strengths and limitations of IMERG can be found at \citet{huffman2023}. IMERG covers the period 2000 onwards, so for consistency both IMERG and ERA5 data are restricted to the period 2000-2015. The blocking and jet metrics we consider (see next section) are very similar in ERA5 whether computed over this period or the longer period 1979-2015, with differences on the order of 5\%. This restriction therefore does not affect our analysis.

The question of how to assess instrument and methodological uncertainty in a data set like IMERG is challenging \citep{huffman2023}. We found that the estimated Gulf Stream precipitation variability (defined in the next section) was around 40\% higher in TRMM/3B42 compared to the more recent IMERG, a far greater difference than what can be explained by sampling variability alone. This suggests that the uncertainty in IMERG could also be large. To highlight this uncertainty, we will at times discuss the differences one would obtain in our analysis if we replaced IMERG with TRMM. TRMM covers 1998-2015, with a native spatial resolution of 0.25°. We restrict TRMM to 2000-2015 also, though note that using 1998-2015 would give almost identical results.

Data is always restricted to the DJF season. All Z500, U850 and SST data are regridded to a regular 2.5° grid. Precipitation data are regridded to a regular 1° grid. When computing SST gradients, SST data are regridded to a regular 0.5° grid, to more accurately distinguish between low and high-resolution models. 

Finally, we note that for a few select metrics, we were unable to obtain the necessary data for one or two models. Thus the sample size of models varies from 135 to 130 across the various figures. Due to the very high sample size this effect is negligible, so we do not draw further attention to this.

\section{Methods.}
\label{sec:methods}

\subsection{Definition of indices and metrics}

Our main goal is to compute correlations capturing intermodel spread of climatological values. In other words, we will compute correlations in scatter plots where each point represents a single model (represented by its climate mean value). In this context, the presence of trends in the underlying timeseries are simply assumed to contribute to the climate mean values, and we make no attempt at untangling forced and unforced contributions to these or characterising differences in trends across models. No trend removal of any kind is therefore performed for the indices/metrics that follow.

\subsubsection{European blocking}
For all datasets studied here, we use the same blocking index as \citet{athanasiadis2022mitigating}, which consists of spatio-temporally averaged instantaneous blocking conditions in the area 30°W-15°E, 45°N-65°N over all DJF seasons in the historical period 1979-2014. This region will be highlighted in a relevant figure.

Instantaneous blocking conditions for each grid point are determined from 500 hPa geopotential height (Z500) fields where two conditions are checked to detect a large reversal in the meridional gradient - a typical signature of blocking \citep{scherrer2006two}. Therefore, we consider that a given grid point with longitude \(\lambda_0\) and latitude \(\phi_0\) is blocked if two conditions, $\text{C}_\text{1}$ and $\text{C}_\text{2}$, are met:
\begin{equation*}
    \begin{aligned}
       &\text{C}_\text{1}: \text{GHGS}(\lambda_0,\phi_0)>0 \\
       &\text{C}_\text{2}: \text{GHGN}(\lambda_0,\phi_0)<-10.\\
    \end{aligned}
\end{equation*}
Here GHGS and GHGN, expressed in meters per latitudinal degree, are 
defined by:
\begin{equation*}
    \begin{aligned}
        \text{GHGS}(\lambda_0,\phi_0)=\frac{Z500(\lambda_0,\phi_0)-
Z500(\lambda_0,\phi_{\text{S}})}{\phi_0-\phi_{\text{S}}}\\
        \text{GHGN}(\lambda_0,\phi_0)=\frac{Z500(\lambda_0,\phi_{\
text{N}})-Z500(\lambda_0,\phi_0)}{\phi_{\text{N}}-\phi_0},\\
    \end{aligned}
\end{equation*}
with \(\phi_{\text{S}}=\phi_0-15^{\circ}\) and \(\phi_{\text{N}}=\phi_0+15^{\circ}\). Consistent with the study area from \citet{athanasiadis2022mitigating}, \(\lambda_0\) here ranges from 30°W to 15°E, and \(\phi_0\) ranges from 45°N to 65°N.

\subsubsection{Northward and southward jet excursions}
Northward jet excursions are defined using the eddy-driven jet latitude index first introduced by \citet{Woollings2010}, which we compute using the simplified methodology of \citet{Parker2019}. Daily zonal winds at 850hPa in the DJF season are first regridded to a common 1 degree grid, and then zonally averaged over the Atlantic domain 0-60W, 15-75N. A 5-day lowpass filter is then applied to remove synoptic variability. For each day, the jet latitude is now defined as the latitude at which the zonally averaged winds are maximum in this smoothed timeseries. The histogram of the jet latitude is trimodal \citep{Woollings2010}, including in the majority of the CMIP6-generation models \citep{DorringtonStrommenFabiano2022}. A northward jet day is then defined, for all models and reanalysis, as one where the jet latitude exceeds 52N, which spans the approximate range of the northernmost peak in a typical trimodal jet latitude histogram. From this we can compute a climatological frequency of northern jet day occurrence over all DJF seasons in the historical period 1979-2014; this will sometimes be referred to as ``NorthDays'' for short. 

It should be emphasised that this is a fairly crude definition, in that it fails to account for split jets, days with insignificant jet activity, or information about the meridional tilt of the jet. However, since the goal here is to understand the impact of Gulf Stream precipitation on ``northward jet excursions'' only in a broad, climate mean-state sense, ignoring such finer scale structure is not unreasonable.


\subsubsection{Gulf Stream precipitation variability}

We follow \citet{schemm2023toward} and define the Gulf Stream region to be the rectangular box 70-50W, 35-45N. This region will be highlighted in relevant figures. As in \citet{schemm2023toward}, precipitation here is to be thought of as a simple proxy for latent heat release by moist diabatic processes. It is possible that more sophisticated measures would provide better results, but as will be seen, this simple proxy already has great explanatory power, so we did not pursue any such increased sophistication. 

As mentioned in the Introduction, we believe an important aspect of our work is the use of a precipitation metric measuring not just the mean precipitation in the Gulf Stream region, but the precipitation variability. This can be motivated \emph{a priori} in at least three ways. Firstly, analysis of the heatflux variability in storm tracks suggests that the variability comes in intermittent `bursts' \citep{messori2013sporadic}, a feature which is elegantly captured in idealized models of jet variability \citep{ambaum2014nonlinear}. Changes to the frequency of occurrence of certain jet configurations or blocking events (e.g., across models) are thus likely going to be more closely related to changes in precipitation variability as opposed to changes in the mean precipitation, which does not see such intermittency. Secondly, as mentioned, \citet{schemm2023toward} showed that the impact of increased resolution in their model was to enhance the northward propagation of the very largest cyclones (those in the upper 99th percentile), with the mean propagation remaining unchanged (cf. Table 1 of \citealt{schemm2023toward}). This suggests that when measuring the links between Gulf Stream precipitation, blocking/jet variability, and resolution, it is likely important to measure higher order moments of the relevant distributions in some way. While Schemm looked at the tails of the cyclone size, we show here that using a measure of precipitation variability appears to achieve similar results, consistent with the conceptual framework described in \citet{ambaum2014nonlinear}. Thirdly, model biases are in general substantially greater when it comes to precipitation variability compared to mean precipitation. This is likely because models are often tuned to achieve a realistic energy budget \citep{hourdin2017art}, and the mean precipitation in models is tightly governed by energetic constraints, with the total amount of precipitation and evaporation being approximately balanced in observations and models on long timescales and large spatial scales \citep{dagan2019analysis}. Using a measure of precipitation variability may therefore be more informative when trying to understand differences in biases across a variety of tuned climate models. 

In this work, we define our measure of Gulf Stream precipitation variability (sometimes referred to as ``GS Precip'' for short) as being the standard deviation of daily precipitation, as computed across all gridpoints in the Gulf Stream region and all days in all DJF seasons in the historical period 1979-2014. In other words, it is the standard deviation of a timeseries of length $\text{time} \times \text{latitude} \times \text{longitude}$ obtained by flattening the three-dimensional $(\text{time}, \text{latitude}, \text{longitude})$ array of daily DJF Gulf Stream precipitation. It therefore measures variability in both time and space. Qualitatively very similar results were obtained if we rather measured the variability using either (a) the maximum monthly precipitation obtained at any of the Gulf Stream gridpoints in any of the December, January and February months (PRmax), or (b) the mean number of ``extreme'' precipitation days (days for which any of the gridpoints in the Gulf Stream region experiences daily precipitation exceeding 40mm) in the DJF season. The standard deviation of daily precipitation was ultimately chosen here because of its simplicity, and the link to the work of \citet{barsugli1998basic} which we will discuss in Section \ref{sec:results}\ref{sec:thermal_damping}. We do not expect that any of the three metrics we considered are optimal measures of precipitation variability for the purposes of understanding blocking/jet variability. Future work should examine this further.

We have also tested the effect of using the mean Gulf Stream precipitation instead, and we discuss the differences in the Results. Due to the skewed and fat-tailed nature of the distribution of precipitation, the mean rainfall in models is correlated with their standard deviation, with a Pearson correlation of around 0.73 using our ensemble. This means that correlation quantities are largely unchanged if precipitation variability is replaced with mean precipitation in our analysis. 

Finally, we found that computing the standard deviation using data on the native model grid, as opposed to regridded data, has a mostly negligible impact. Our results are therefore not sensitive to this choice. This is plausibly related to the fact that the size of the Gulf Stream domain is small relative to the typical grid spacing of the models considered.

\subsubsection{Atmospheric horizontal resolution}

To measure model resolution, we focus on atmospheric horizontal resolution, because oceanic resolution is strongly correlated with atmospheric resolution across models (i.e., modelling centres tend to increase both in tandem) and atmospheric resolution varies more finely across models. We use the nominal grid spacing at the equator as our measure of nominal atmospheric resolution. This nominal resolution of the models was obtained by reading the relevant model implementation paper for all the models, converting from spectral resolution to grid spacing where necessary. The results are summarised in Supporting Tables S1, S2 and S3. We note that in a few limited cases it was not always possible to find a written quote about a specific model version. In these cases we assumed the resolution matched that of the most closely matching model version. 

\subsection{Estimating internal variability of blocking and jet variability}

The North Atlantic jet and European blocking are known to exhibit notable decadal variability in both observations and reanalysis \citep{hurrell1997decadal, woollings2018daily, DorringtonStrommenFabiano2022}. This can complicate questions around model biases, since models may show discrepancies from observations due to inhabiting different modes of unforced (``internal'') decadal variability from the real world, independently of any failure to resolve the physical processes correctly. It is thus important to include some measure of this internal variability in observational estimates of European blocking and northward jet excursions. We do this here by a simple bootstrap procedure, whereby we resample random ERA5 years with replacement to generate 1000 timeseries, and define the magnitude of internal variability to be twice the standard deviation of the resulting distribution. Estimating internal variability by rather selecting random chunks of years (of the correct size) from 1900-2010 using the 20th century reanalysis product ERA20C \citep{Poli2013} yielded similar results.

Note that the analysis of \citet{woollings2018daily} suggests that recent observed blocking frequencies are if anything on the lower end of what might be expected from decadal variability (cf. Figure 11 therein). The apparent underestimation of European blocking and northern jet days in models is thus unlikely to just be an artefact of having compared models and observations over a period where the real world happened to be in a phase of decadal variability involving above-average blocking.

\subsection{Statistical analysis}

Because of the non-Gaussianity of the distribution of precipitation, correlations are always computed as Spearman rank correlations, using the \texttt{spearmanr} function from the python package \texttt{scipy} \citep{2020SciPy-NMeth}. However, despite non-Gaussianity of some of the individual distributions considered, the observed relationships between variables are in almost all cases approximately linear, and so using Pearson correlations gave virtually identical results in almost all cases. The one exception here is the observed link between precipitation variability and atmospheric horizontal resolution, which shows evidence of non-linearity. In this case the Pearson correlation appears to underestimate the observed link, with the Spearman correlation being higher.

To estimate statistical significance of any computed correlations, we almost always use bootstrap resampling, in order to avoid the use of standard tests that assume Gaussianity. Concretely, given samples $x$ and $y$ and a null hypothesis of there being no relationship, we resample $x$ and $y$ randomly with replacement 10,000 times and correlate the resulting random samples to obtain a distribution of correlations for the null hypothesis. Given our typical sample size of $N=135$, this method always produced a 95\% confidence interval for the null hypothesis of around $\pm 0.19$, and a 99\% interval of around $\pm 0.22$. We thus interpret correlations exceeding $0.2$ in magnitude as unlikely to be random. The exception to our use of bootstrap resampling is when testing for the effect of mediation, where we use an explicit Sobel test \citep{baron1986moderator}.


A weakness of our statistical analysis is that we deliberately do not distinguish between simulations performed by different models and different ensemble members simulated using a single model. This is done in order to increase the sample size as much as possible, and to ensure any links uncovered are robust to the large internal variability ensembles using single models can exhibit \citep{scaife2018signal}. The drawback is that ensemble size varies a lot between the models, meaning that estimates may sometimes be affected by, for example, the presence of a particularly large single-model ensemble at a particular horizontal resolution. Visual inspection of the relevant scatter plots suggests that this effect is probably small, and we confirmed that correlations do not substantially change if only a single ensemble member is retained for each model. It is worth remarking that different climate models are not generally independent of each other either, and this can inherently bias any multimodel analysis, irrespective of the number of ensemble members retained \citep{knutti2013climate, annan2017meaning, boe2018interdependency}.

\section{Results}
\label{sec:results}

\subsection{Model biases in the mean and variability of key variables}
\label{sec:model_biases}

\begin{figure}[ht!]
\centering
  \noindent\includegraphics[width=35pc,angle=0]{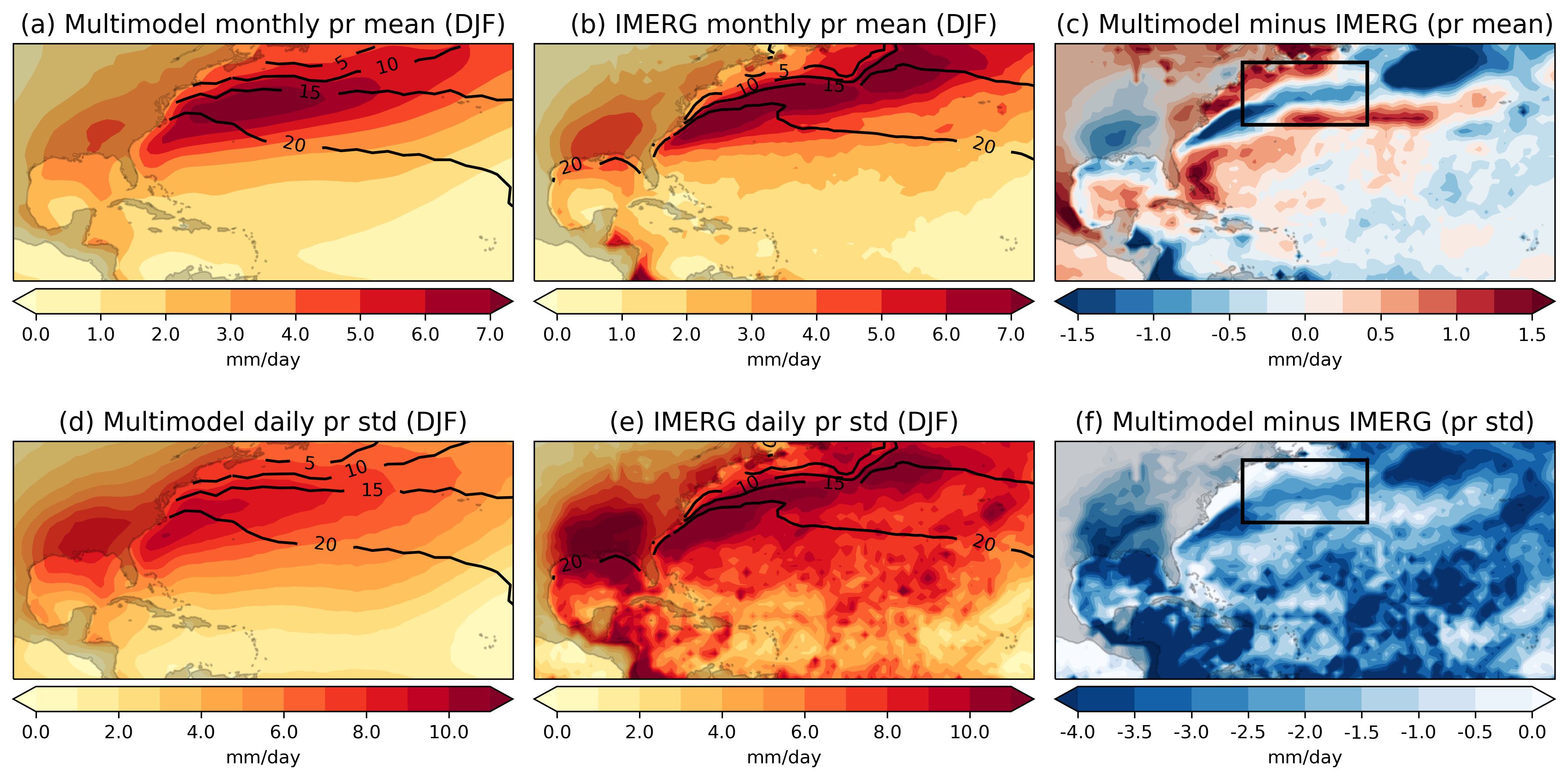}\\
  \caption{In (a): the multimodel mean monthly precipitation (filled contours), along with the multimodel mean SSTs (line contours), at each gridpoint. In (b): the mean monthly IMERG precipitation (filled contours), along with the ERA5 mean SSTs (line contours). In (c), the difference (a) minus (b). In (d), the multimodel mean daily standard deviation of precipitation (filled contours), and multimodel mean SSTs (line contours). In (e) the standard deviation of daily IMERG precipitation. In (f) the difference (d) minus (e). The season is always restricted to DJF. The period covered is 1979-2014 for model data and 2000-2015 for observations/reanalysis. In (c) and (f) the Gulf Stream region is highlighted with a black box.}\label{fig:precip_biases}
\end{figure}

\begin{figure}[ht!]
\centering
  \noindent\includegraphics[width=35pc,angle=0]{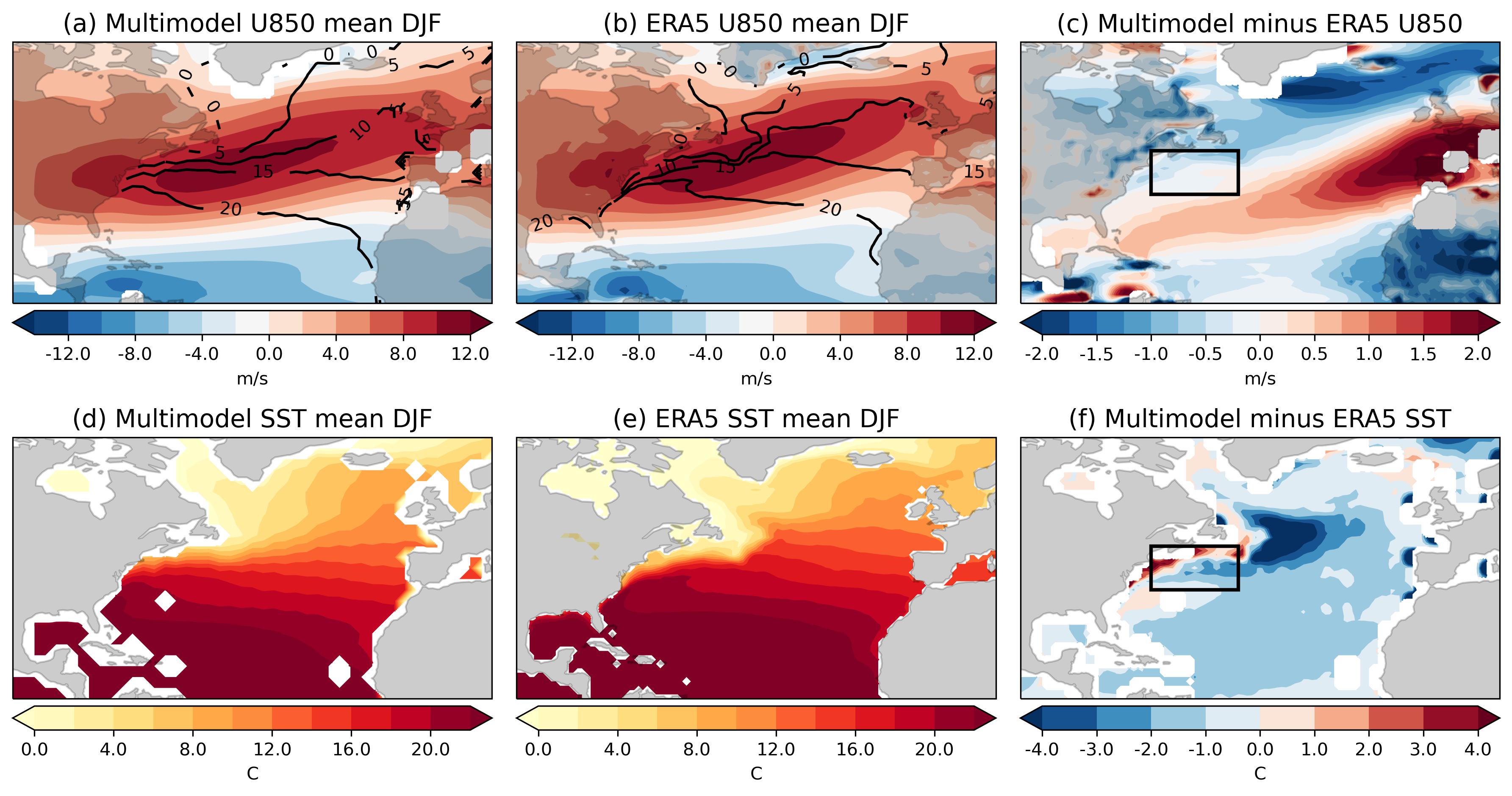}\\
  \caption{In (a): the multimodel mean monthly 850hPa zonal winds (filled contours), along with the multimodel mean SSTs (line contours), at each gridpoint. In (b): the mean monthly ERA5 850hPa zonal winds (filled contours), along with the ERA5 mean SSTs (line contours). In (c), the difference (a) minus (b). In (d), the multimodel mean SSTs (filled contours). In (e) the mean ERA5 SSTs. In (f) the difference (d) minus (e). The season is always restricted to DJF. The period covered is 1979-2014 in all cases. In (c) and (f) the Gulf Stream region is highlighted with a black box. For a discussion on the occasional missing values in model data, see Methods.}\label{fig:sst_biases}. 
\end{figure}

For context, we start by visualising biases of key quantities in the multimodel ensemble. Note that similar multimodel analysis has been carried out in several previous papers, including for precipitation \citep{yazdandoost2021evaluation, abdelmoaty2021biases}, SSTs \citep{zhang2023understanding} and zonal winds \citep{harvey2020response, athanasiadis2022mitigating}. Our analysis is included here anyway to make the present paper self-contained.

Figures \ref{fig:precip_biases}(a)-(c) shows the mean monthly precipitation for the multimodel mean, IMERG and the multimodel bias (multimodel mean minus IMERG). The SST contours for the multimodel mean and ERA5 are shown as well. In both models and observations, there is a strong band of precipitation following the Gulf Stream, as expected \citep{minobe2008influence}. However, the band is generally more diffuse in models, consistent with the failure of models to reproduce the early Gulf Stream separation, and consequent sharp SST gradients, observed in the real world. Comparing Figures \ref{fig:precip_biases}(b) and (c), one can estimate that, in the Gulf Stream region (highlighted with a black box), the mean bias amounts to an underestimation of total precipitation of roughly 15\%. Figures \ref{fig:precip_biases} (d)-(f) show the analogous plots but for the standard deviation of daily precipitation. Comparing Figures \ref{fig:precip_biases}(e) and (f) suggests the typical underestimation of Gulf Stream precipitation variability in models is roughly 35\%. The equivalent plot using TRMM rather than IMERG is given as Figure S1 in the Supplemental Material. TRMM places the peak precipitation more clearly in the Gulf Stream box, unlike IMERG, which produces peaks further up and downstream. When compared to TRMM, the multimodel underestimation of precipitation variability is therefore much higher, at around 60\%. It is unclear to the authors whether TRMM or IMERG gives the more realistic estimate.

Figures \ref{fig:sst_biases}(a)-(c) (top row) show the mean 850hPa zonal winds (U850) for models, ERA5 and the model bias, while (d)-(f) (bottom row) shows the analogous plot but for SSTs. The zonal wind biases show the ``too zonal, too equatorwards'' bias emphasised in \citet{schemm2023toward}. Figure \ref{fig:sst_biases}(c) also highlights a strong negative bias emerging from the southern tip of Greenland. This is the region highlighted when doing composite analysis of the northern jet latitude events \citep{Woollings2010}, and this negative zonal wind bias is likely therefore partially reflecting model biases in such northern jet events. It is likely also reflecting European blocking biases \citep{o2017gulf, DorringtonStrommenFabiano2022}. Figure \ref{fig:sst_biases}(f) clearly highlights two classic model biases: the reduced SST gradient in the Gulf Stream separation region, and the large cold bias in the central sub-polar North Atlantic. As explained in the Introduction, these biases are closely linked to the failure of models to simulate the early Gulf Stream separation and ``northwest corner''.

\subsection{The link between Gulf Stream precipitation, European blocking and northward jet excursions}
\label{sec:precip_link}

\begin{figure}[ht!]
\centering
  \noindent\includegraphics[width=18pc,angle=0]{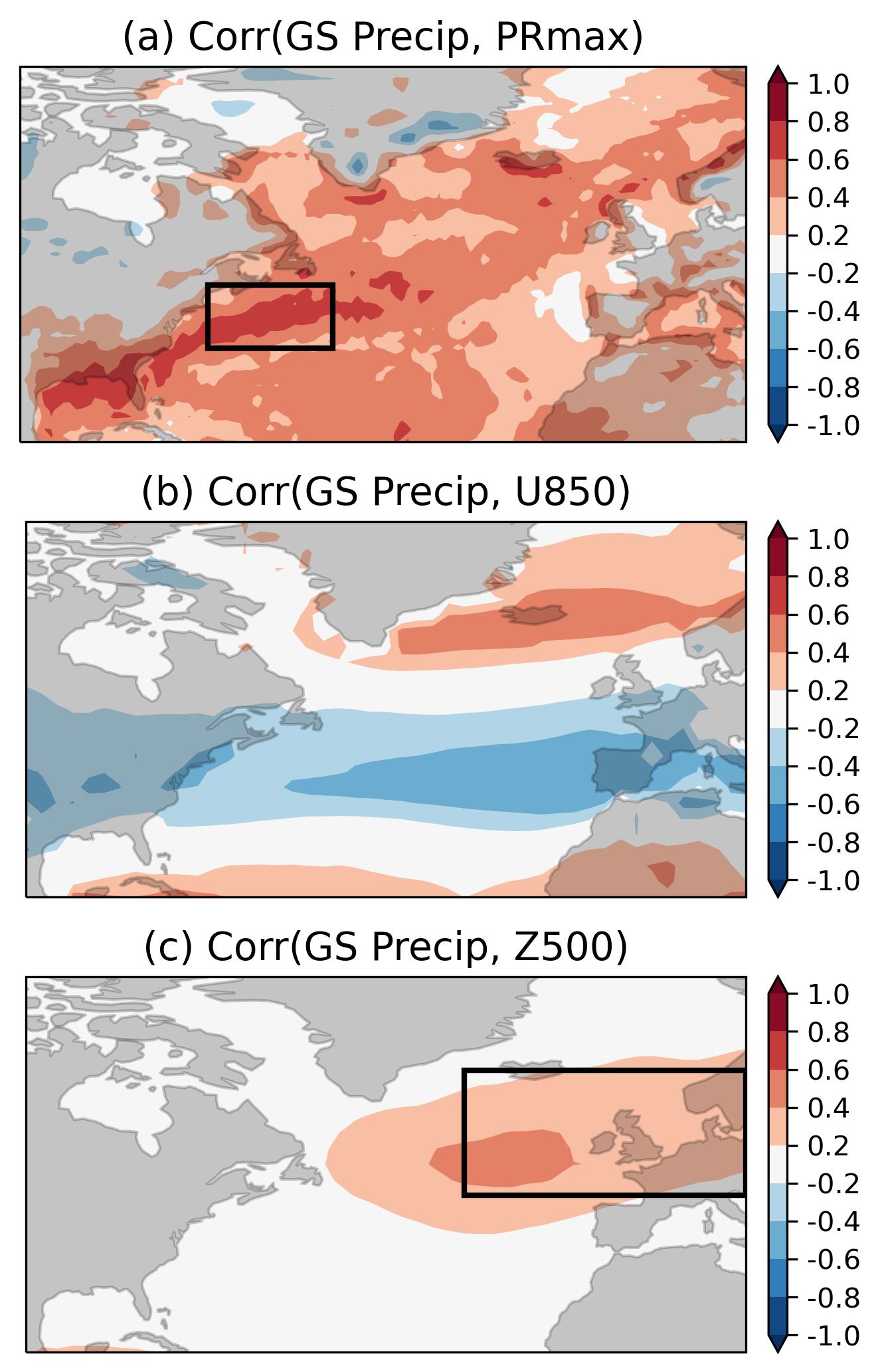}\\
  \caption{Correlations between GS Precip and the local (a) climatological maximum monthly precipitation (\(\text{PR}_\text{max}\)) obtained across any of the winter months; (b) climatological DJF U850; (c) climatological DJF Z500. Correlations are computed across the multimodel ensemble. The period considered is always 1979-2014. Correlations outside the zero contour ($\pm 0.2$) are statistically significant ($p<0.05$). In (a), the Gulf Stream region is highlighted with a black box, and in (c) the region used to define European blocking is similarly highlighted.}\label{fig:spatial_corrs}
\end{figure}

\begin{figure}[ht!]
\centering
  \noindent\includegraphics[width=22pc,angle=0]{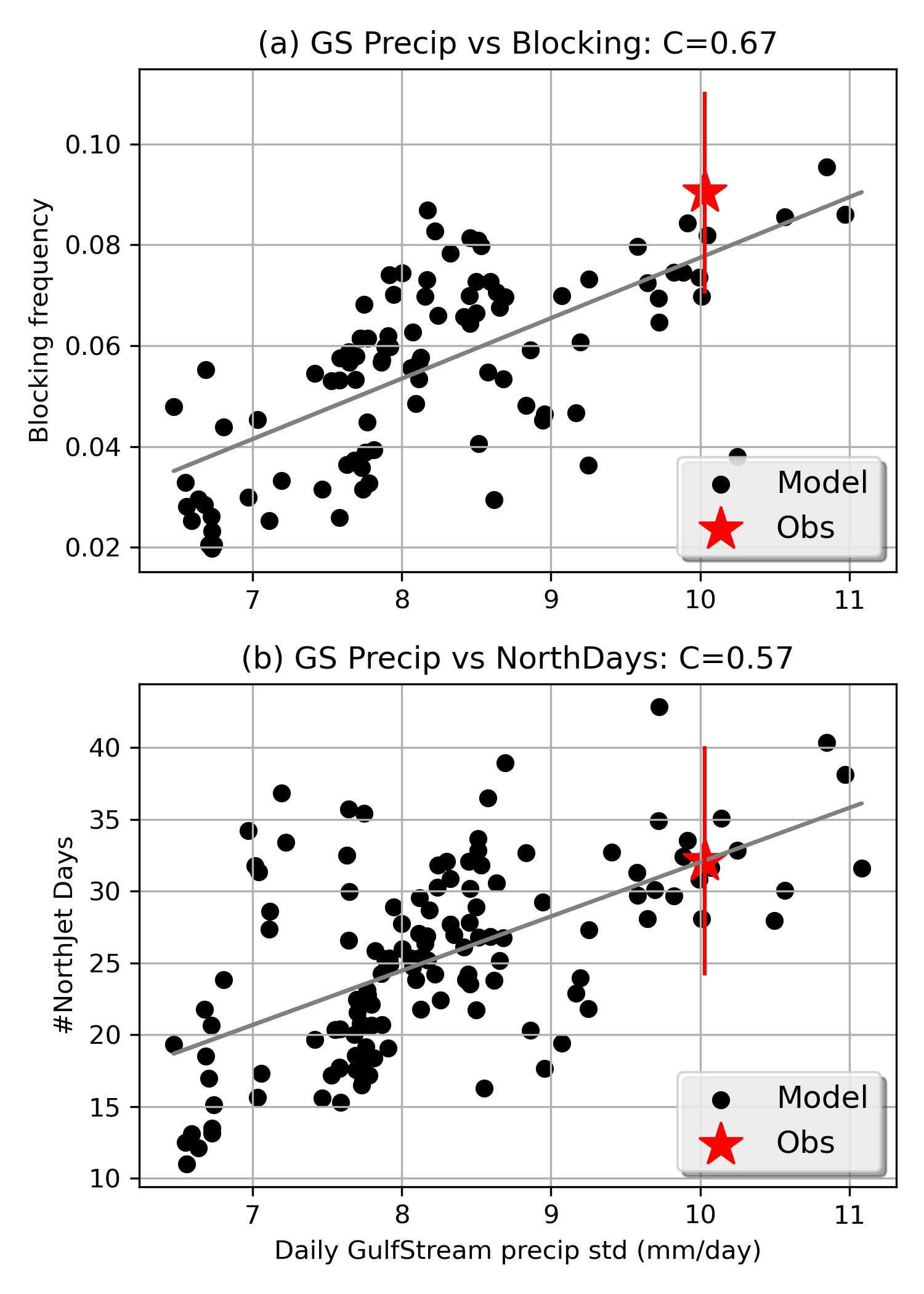}\\
  \caption{Scatter plots of Gulf Stream precipitation variability (GS Precip) versus (a) European blocking frequency and (b) northern jet days. Black dots are models, while the red star is observations (IMERG/ERA5). The red lines indicate our measure of internal variability in observations. The grey lines in each panel is a linear fit to the black dots. The value of $C$ in each title is the correlation measured using model data. Model data covers the period 1979-2014, while observations cover 2000-2015.}\label{fig:scatter_precip}
\end{figure}

We now examine the link between Gulf Stream precipitation variability and European blocking/northern jets. We compute correlations across the multimodel ensemble between a model's GS Precip and the climatological mean of PRmax, U850 and Z500 at each spatial location. This is shown in Figure \ref{fig:spatial_corrs}. Thus, if a correlation at a particular gridpoint is positive (negative), it means that models with greater Gulf Stream precipitation variability tend to have a greater (lower) climatological mean PRmax, U850 or Z500 at that gridpoint. Figure \ref{fig:spatial_corrs}(a) shows that models with greater GS Precip also experience enhanced precipitation maxima broadly throughout the North Atlantic domain, consistent with both the model fundamentally simulating precipitation better and also with dynamical downstream effects via the storm track. Figure \ref{fig:spatial_corrs}(b) shows that greater GS Precip is also associated with more northerly zonal winds, compensating the negative bias seen in Figure \ref{fig:precip_biases}(c). As discussed last subsection, this is strongly suggestive of more northern jet days, something we confirm shortly. Finally, Figure \ref{fig:spatial_corrs}(c) shows that greater GS Precip is associated with a positive Z500 anomaly over the sub-polar North Atlantic, stretching into Europe. The black box here highlights the region used to define the European blocking index; Figure \ref{fig:spatial_corrs}(c) is thus suggestive of a link between GS Precip and blocking.

These impressions are confirmed in the scatter plots shown in Figure \ref{fig:scatter_precip}. Figure \ref{fig:scatter_precip}(a) shows a visually clear link between GS Precip and European blocking in models: the correlation is $0.67$, with approximately 45\% of the spread in European blocking frequencies being accounted for by GS Precip. For NorthDays, the correlation is $0.57$, corresponding to around 33\% of variance explained. The somewhat smaller correlation for NorthDays may be because the impact of GS Precip on the jet consists of both an increase in the northern jet latitude regime \emph{and} a reduction in the central jet regime (see Figure \ref{fig:spatial_corrs}(b)), meaning the full impact on the jet latitude variability is not visible solely in the NorthDays metric. It should also be recalled that while European blocking events are most typically associated with a northern jet, the two metrics are relatively decoupled from each other on daily timescales, with the jet often found at central and southern latitudes when blocking is ongoing \citep{Woollings2010}.

While Figure \ref{fig:scatter_precip} gives the impression that some models are able to achieve realistic levels of precipitation variability, a very different impression would have been given if we had used TRMM in place of IMERG. We comment more on this in the following subsection. A different impression would also be given if we had used the mean Gulf Stream precipitation, rather than a measure of variability. Supplemental Figure S2 shows the equivalent of Figure \ref{fig:scatter_precip} if mean precipitation is used. This shows that while the multimodel mean is biased too low, a large proportion of models clearly exceed the mean precipitation of IMERG, in clear contrast to the systematic underestimation of European blocking. If a measure of variability or extremes is used instead, IMERG and TRMM will both appear at the upper end of what models are able to simulate, in better agreement with the situation for European blocking.

Finally, we note that if we only retain a single ensemble member for every model, the correlations between GS Precip and European blocking/NorthDays are around $0.5$ for both. Thus the link highlighted here is not an artefact of a few individual models with many ensemble members.

\subsection{The effect of atmospheric horizontal resolution}
\label{sec:resolution_impact}

\begin{figure}[ht!]
\centering
  \noindent\includegraphics[width=22pc,angle=0]{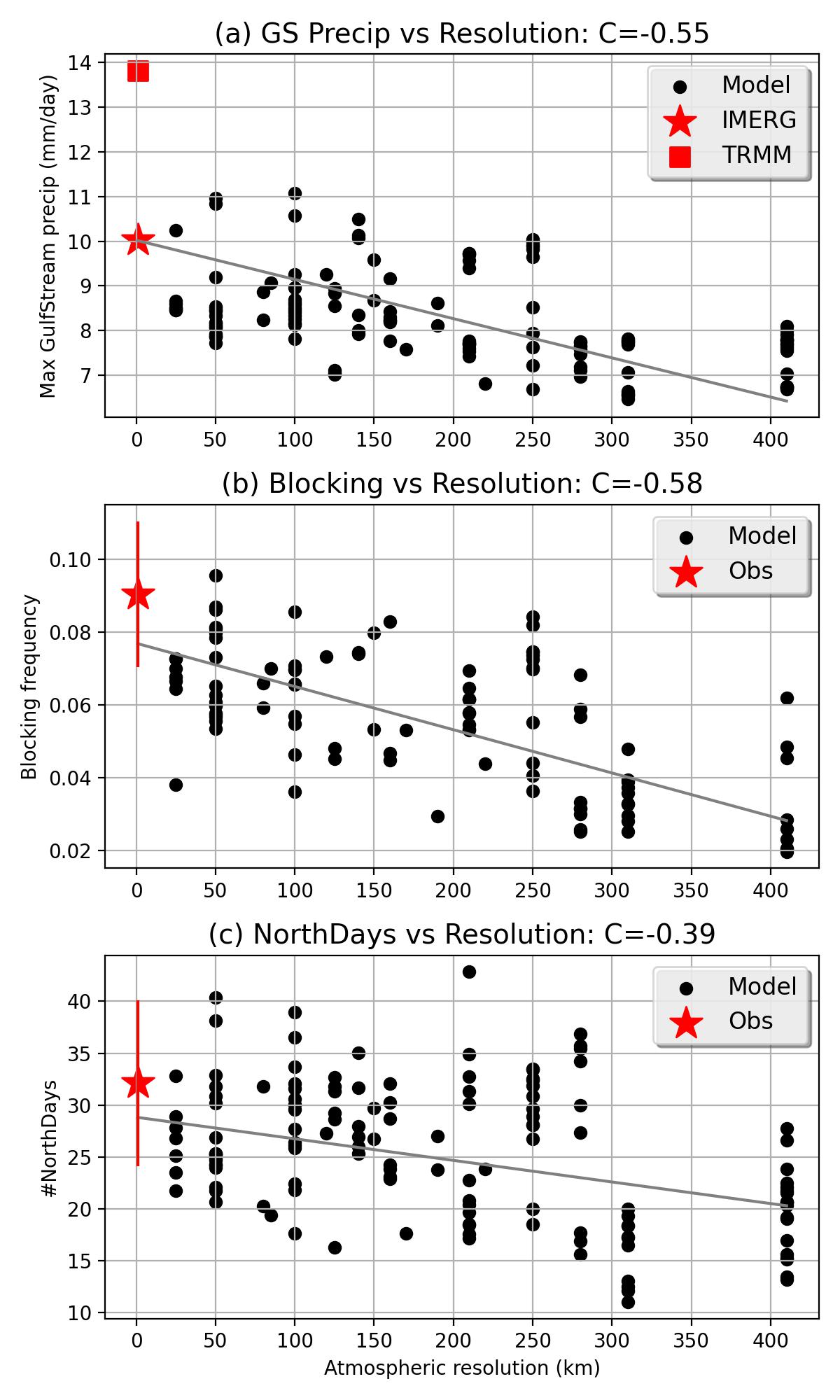}\\
  \caption{Scatter plots of atmospheric horizontal resolution (grid spacing at the equator in km) versus (a) Gulf Stream precipitation variability, (b) European blocking frequency, and (c) northern jet day occurrence. Black dots are models, while the red star is observations (IMERG/ERA5). The red lines in (b) and (c) indicate our measure of internal variability in observations. The grey lines in each panel is a linear fit to the black dots. The value of $C$ in each title is the correlation measured using model data. Model data covers the period 1979-2014, while observations cover 2000-2015.}\label{fig:scatter_resolution}
\end{figure}

In the previous section we have seen that Gulf Stream precipitation variability is strongly linked to European blocking frequency and northern jet excursions. We now examine how all these three factors are linked to atmospheric horizontal resolution.

Figure \ref{fig:scatter_resolution} shows scatter plots of horizontal resolution versus (a) GS Precip, (b) European blocking frequencies, and (c) northern jet days. Significant correlations are found in all cases, with the highest correlation being that of $-0.58$ between resolution and blocking. This correlation is identical to that reported in \citet{davini2020cmip3} between resolution and a closely related blocking metric\footnote{\citet{davini2020cmip3} compute DJFM blocking frequencies over a central/eastern European region which partially overlaps with the region we consider, over the period 1961-2000.} using a similarly large multimodel dataset: see their Figure 3(b). The link between resolution and European blocking therefore appears to be particularly robust. As in the previous section, the correlation for NorthDays is smaller.

Figure \ref{fig:scatter_resolution}(a) confirms the expected association between enhanced precipitation variability and higher atmospheric resolution, noted in a variety of past modelling studies \citep{kopparla2013improved, bador2020impact, strandberg2021importance}, including ones focused on the Gulf Stream region in particular \citep{scher2017resolution}. IMERG gives the clear impression that a nominal atmospheric resolution of around 25km is sufficient to simulate realistic precipitation variability; note that the intersection between the red star and the grey line in (a) is not by construction. Strikingly however, TRMM is considerably higher than the 25km models, and a major step change would need to take place at higher resolutions in order to bring model variability in line with TRMM. In fact, the possibility that such a major step change may occur once a threshold resolution has been exceeded has been hypothesised before \citep{slingo2022ambitious}. Previous studies, like \citet{scher2017resolution} and \citet{strandberg2021importance}, have shown that the impact of atmospheric resolution on the rainfall distribution is asymmetric, with little/no change to low magnitude events and larger changes to extreme events. Such asymmetries, likely related to the increasingly accurate ability of high-resolution models to resolve individual storm systems, are often associated with non-linearities, which a step change would be an example of. A non-linear relationship can also be motivated by a simple conceptual model. Precipitation simulated by a climate model in a gridbox is, roughly speaking, obtained as an average over individual systems taking place therein. One can therefore imagine a model's precipitation as a spatially smoothed version of real world precipitation. To assess how such smoothing affects the standard deviation, one can, for example, apply a uniform or Gaussian filter to fixed samples from a Pareto distribution, which are commonly used as a statistical model for precipitation \citep{cavanaugh2015probability}. This shows that the standard deviation drops off sharply as the window size of the filter increases (i.e., as the samples are smoothed more), before flatlining as the effect of additional smoothing begins to saturate: see Supplemental Figure S3. Thus a steep rise in variability as models exceed 25km resolution cannot be ruled out.

Interestingly, there is some indication of non-linearity in Figure \ref{fig:scatter_resolution}(a), since the Spearman correlation of $-0.55$ is greater in magnitude than the corresponding Pearson correlation of $-0.45$, indicative of a monotonic relationship beyond the purely linear. However, we do not pursue this line of thought further and do not make any robust claims. We only discuss these potential non-linearities because of the relevance these would have for model development. Our actual results are not affected either way.

Based on the results seen so far, we hypothesise that the pathway from ``increased horizontal resolution'' to ``reduction in blocking/jet biases'' is mediated, at least in part, by ``improved Gulf Stream precipitation variability''. The standard test for mediation \citep{baron1986moderator}, in the context of European blocking specifically, is therefore as follows:
\begin{itemize}
    \item[1.] Regress resolution against European blocking and check that there is a significant correlation.
    \item[2.] Regress GS Precip against European blocking and check that there is a significant correlation.
    \item[3.] Do a multiple regression of resolution \emph{and} precipitation against European blocking and assess if the regression coefficient corresponding to resolution is significantly lower than that obtained in step 1.
\end{itemize}
\noindent The procedure is analogous for NorthDays. Steps 1 and 2 here have been affirmed in the positive with Figures \ref{fig:scatter_precip} and \ref{fig:scatter_resolution}. We carried out step 3 separately: the regression coefficient associated to resolution in step 1 was estimated as $-0.73$ (using timeseries that are normalised to have mean zero and standard deviation 1), and this halves to $-0.36$ in step 3. This reduction was estimated as statistically significant ($p<0.05$) using the Aroian version of the Sobel test, described in \citet{baron1986moderator}. Thus, the data we have analysed here supports our hypothesis.

\subsection{Precipitation variability in AMIP models}
\label{sec:thermal_damping}

We will now look briefly at how Gulf Stream precipitation variability differs between AMIP and coupled models. We do this here by referring to the classic work of \citet{barsugli1998basic} (hereafter BB98). The implications this analysis has for assessing the relative roles of atmospheric and oceanic resolutions will be considered in the Discussion.

\begin{figure}[ht!]
\centering
  \noindent\includegraphics[width=22pc,angle=0]{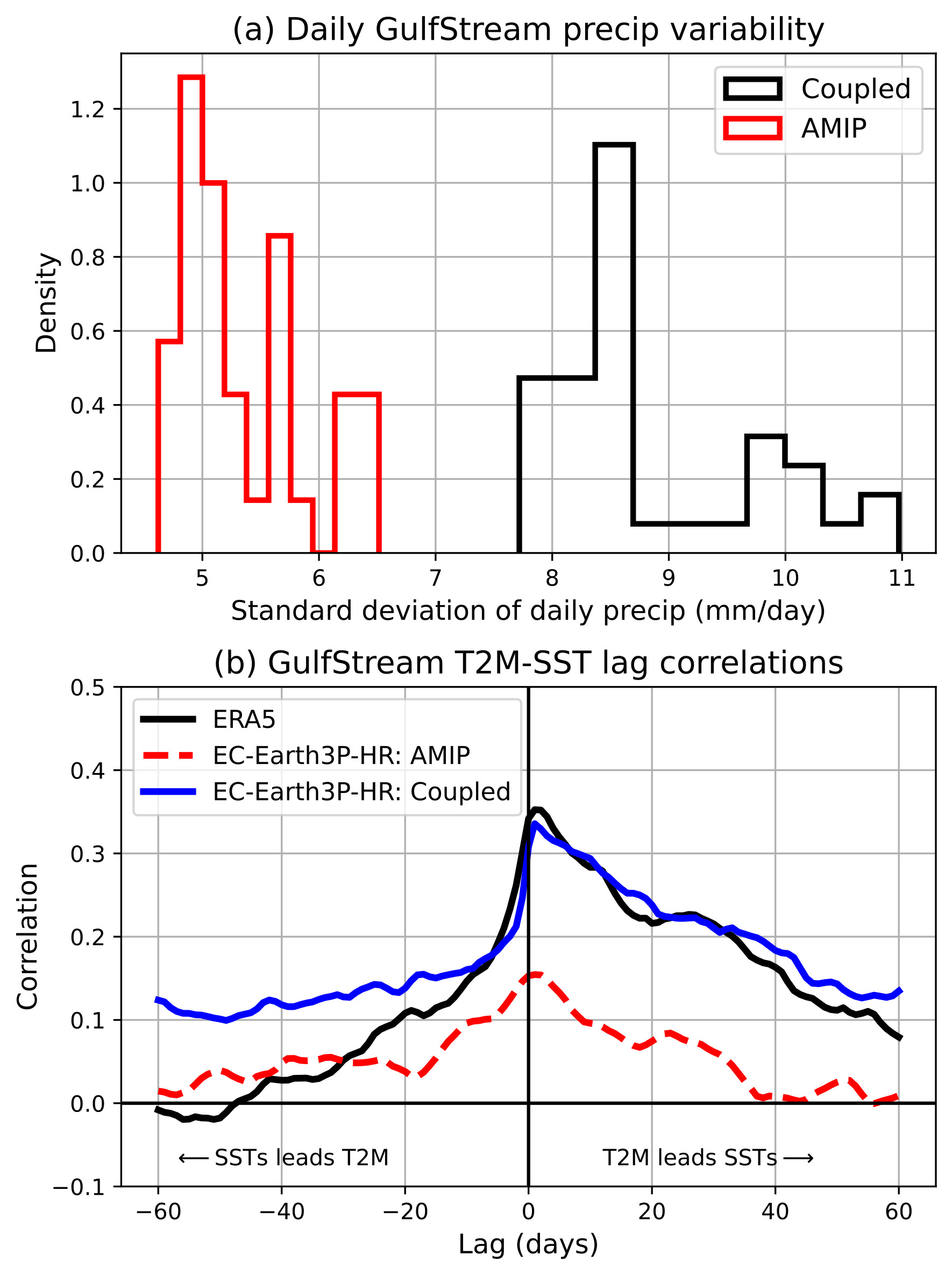}\\
  \caption{In (a): histograms of the daily standard deviation of Gulf Stream precipitation for the coupled PRIMAVERA/HighResMIP models (black) and the counterpart AMIP models (red). In (b): lead-lag correlations of daily Gulf Stream SST vs daily Gulf Stream T2M for ERA5 (black), the coupled version of EC-Earth3P-HR (blue) and the AMIP version of EC-Earth3P-HR (red stipled). Negative lags correspond to SSTs leading T2M, and positive lags to T2M leading SSTs.}\label{fig:amip_vs_ocean}
\end{figure}

BB98 constructed a simple pair of stochastic ordinary differential equations to model air-sea coupling, and used it to explain observed differences between coupled models and AMIP-style models. It is well known that heatfluxes are unrealistic in AMIP models, due to the ocean being an infinite source/sink of energy, and this is explained by the BB98 model. What seems to be less well known is that AMIP models have systematically reduced daily surface temperature variability compared to coupled models, due to what BB98 describe as the excessive thermal damping taking place in AMIP models. A self-contained presentation of the BB98 model and how it accounts for these differences between AMIP and coupled models is given in the Appendix. 

The reason this is relevant is because the reduced variability of daily surface temperatures results in a similarly reduced variability of daily precipitation, due to the effect of temperature on convective processes, such as condensation. Figure \ref{fig:amip_vs_ocean}(a) shows the standard deviations of daily Gulf Stream precipitation for the AMIP simulations of HighResMIP (red histogram) and the corresponding set of coupled simulations (black histogram). It can be seen that even the highest resolution AMIP model experiences variability significantly lower than the lowest resolution coupled model, with the variability being around 40\% lower on average in the AMIP models. BB98 show that this is due to the missing coupling, as measured by the lead/lag-correlations of SSTs versus surface temperature (T2M). An example of this using the EC-Earth3P-HR model is shown in Figure \ref{fig:amip_vs_ocean}(b), reproducing what is expected from the analysis of BB98.

\citet{schemm2023toward} found in his AMIP simulations that the main impact was on the most intense cyclones only. It would be interesting to explore if this reflects the thermal damping effect, and whether or not a more widespread impact would have been seen if his simulations included air-sea coupling.

\section{Discussion}
\label{sec:discussion}

\subsection{Mechanisms and causality}
\label{sec:mechanisms}

The mechanism suggested by Schemm's work is that the improved representation of diabatic processes results in more cyclones undergoing explosive growth and propagating further poleward. Testing this here would require Lagrangian cyclone tracking in the 135 model simulations we have made use of. While we believe much could be learned by carrying out such tracking in model data and comparing to observations, we did not do this here due to the intense computational cost involved.

\begin{figure}[ht!]
\centering
  \noindent\includegraphics[width=30pc,angle=0]{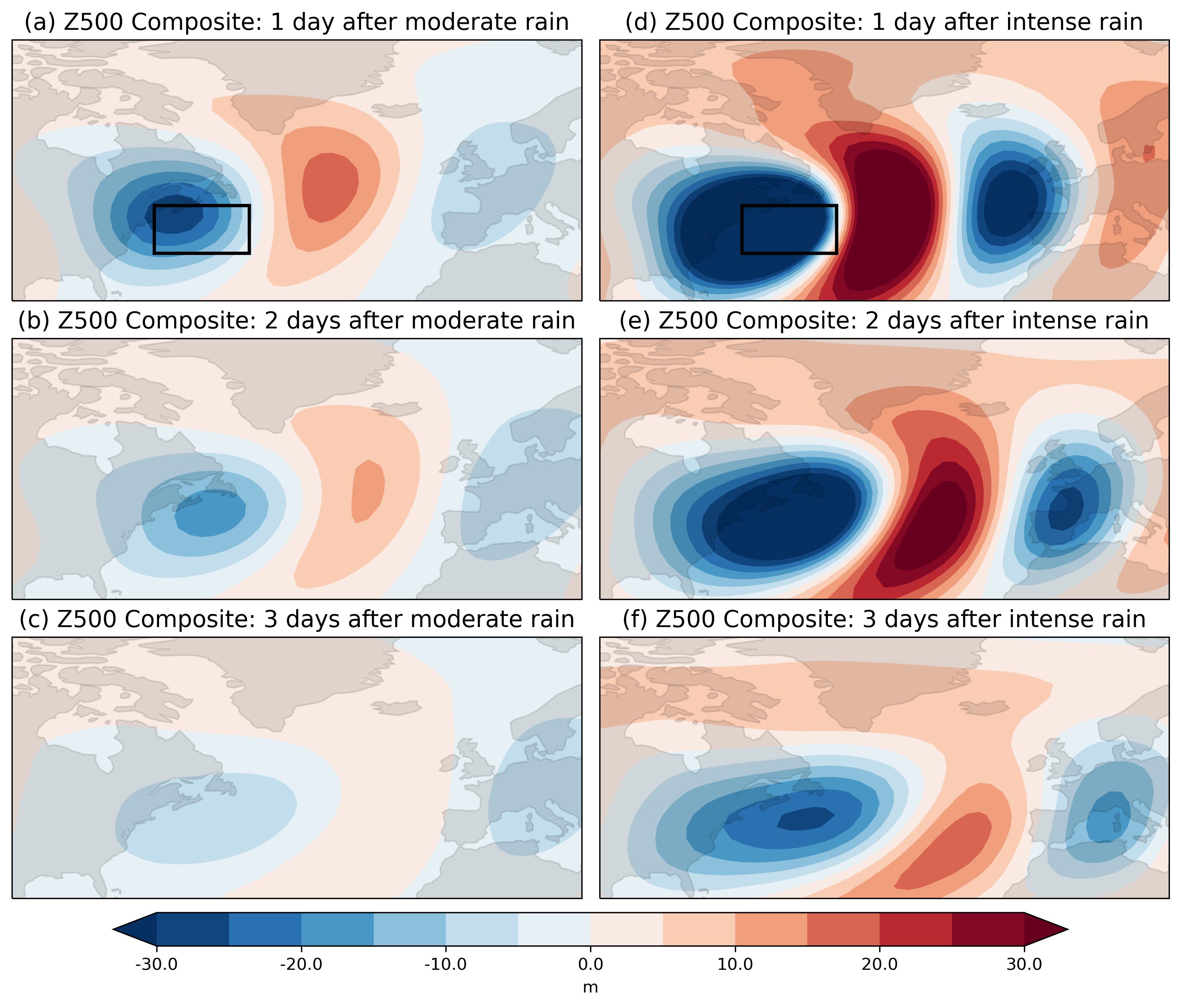}\\
  \caption{Anomalous Z500 composites, across all model data, in the 3 days following a day with moderate Gulf Stream precipitation (left column, (a), (b) and (c)), and intense Gulf Stream precipitation (right column, (d), (e) and (f)). Moderate precipitation days are defined as days where the area-averaged Gulf Stream precipitation is between 5 and 10mm, and intense precipitation days as ones where the averaged precipitation exceeds 15 mm. The Gulf Stream box is highlighted in (a) and (d). Model data is restricted to 1979-2014.}\label{fig:composites}
\end{figure}

However, as a basic sanity check, we computed composites (across all model data) of Z500 anomalies in the aftermath of a day where the area-averaged Gulf Stream precipitation was moderate (between 5 and 10mm accumulated over the day) or intense (more than 15mm). The results are shown in Figure \ref{fig:composites}. This shows that in the model ensemble, intense precipitation days are followed by a downstream wave-train which is much stronger than that following moderate precipitation days, with some suggestion of further poleward migration as well. These composites were found to be qualitatively similar across sub-ensembles and individual models (not shown). While they show a low over Europe on the days considered, rather than the high expected during European blocking, it should be kept in mind that (a) European blocking is a relatively rare occurrence, and (b) the composites are across models that systematically underestimate the occurrence of blocking. These composites will therefore not generally be averaging over days where European blocking is taking place. However, the composites suggest that, all else being equal, models that experience more days with heavy precipitation would be expected to generate more and stronger cyclones, which under the right conditions can then transport negative potential vorticity from the air-sea interface to the upper troposphere and contribute to a block \citep{vanniere2016potential, mathews2024oceanic}. Future work should look more closely at the transient evolution of blocks in multi-model ensembles and the contribution of precipitation, as has been done for reanalysis data \citep{marcheggiani2023diabatic, wenta2024linking, mathews2024oceanic}.

Given the free-running nature of the models we examined, it cannot be taken for granted that the correlations we have highlighted are indicative of a causal pathway from precipitation variability to European blocking/northern jet excursions. The experiments of \citet{mathews2024gulfstream} demonstrate that causality certainly \emph{can} go in this direction, and our test for mediation also corroborates this causality. Could the causality go the other way as well? The most obvious way to imagine this being the case is if individual episodes of European blocking promote enhanced precipitation in the Gulf Stream region. However, previous studies on the storm track and blocking suggest the opposite should be the case, since (1) the generation of cyclones in the Gulf Stream region act to reduce the excess baroclinicity, thus terminating the burst of storm track activity \citep{ambaum2014nonlinear}; (2) the block acts to deplete the excess ocean thermal heat content by feeding on it, thereby reducing the potential for strong air-sea fluxes \citep{mathews2024oceanic}. Thus, on the face of it, European blocking and northern jet excursions are more likely to lead to a suppression of Gulf Stream precipitation variability than an enhancement.

\subsection{Alternative hypotheses: the role of SST biases}
\label{sec:alternative_hypotheses}

The most obvious alternative explanation for the correlations identified in Figures \ref{fig:scatter_precip} and \ref{fig:scatter_resolution} is that there is some other process or component of the climate system which modulates both blocking/jet and precipitation variability.

In terms of processes, one natural possibility would be transient eddy feedbacks, since one could imagine a stronger feedback (a) invigorating storm systems in the Gulf Stream, thereby enhancing precipitation variability, and (b) also allowing cyclones to persist for longer and thereby amplify blocking anticyclones. However, \citet{DorringtonStrommenFabiano2022} showed that eddy feedbacks do not systematically increase with horizontal resolution in the multimodel ensemble considered here, unlike blocking/NorthDay frequency and precipitation variability. Variations in eddy feedback strength therefore would not be able to explain the correlations we find here. Another possibility is that atmospheric resolution improves the model's ability to simulate the steep vertical ascent over the Gulf Stream associated with fronts, which could also affect jet variability, as suggested in forthcoming work by Robert Wills et al. We consider this an important avenue for future work which could clearly complement the analysis of our paper.

\begin{figure}[ht!]
\centering
  \noindent\includegraphics[width=35pc,angle=0]{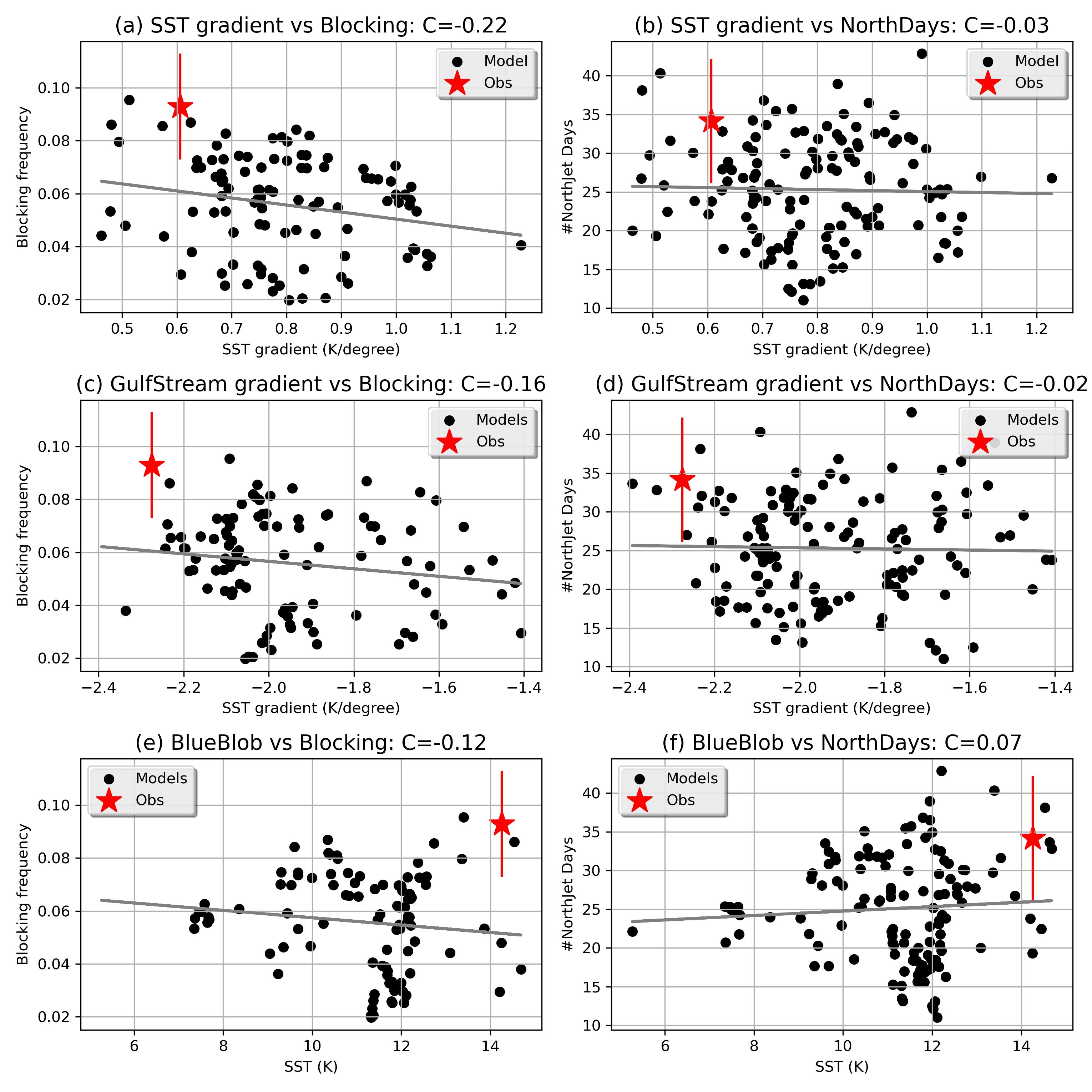}\\
  \caption{Scatter plots of the SST gradient in the central North Atlantic versus (a) European blocking frequency and (b) northern jet day occurrences (NorthDay). In (c) and (d) the same but for the Gulf Stream SST gradient. In (e) and (f) the same but for the North Atlantic cold SST bias (``blue blob of death''). Black dots are models, while the red star is observations (IMERG/ERA5). The red lines indicate our measure of internal variability in observations. The value of $C$ in each title is the correlation measured using all the model data. Model data covers the period 1979-2014, while observations cover 2000-2015. The SST gradient metric is directly adapted from that used in \citet{athanasiadis2022mitigating}.}\label{fig:sst_scatter}
\end{figure}

In terms of other components of the climate system, the most obvious ones are SST biases, given the consistent attention given to such biases in prior literature. To test if SST biases can explain our results, we tested three different SST metrics. Firstly, the SST gradient in the Gulf Stream extension region (45-5W, 40–50N): this is the bias that \citet{athanasiadis2022mitigating} argued is responsible for biases in European blocking/NorthDays. Secondly, the SST gradient in the Gulf Stream region we have worked with here (70-50W, 35-45N): gradients in this region have been argued to influence blocking/NorthDays in \citet{o2016influence} and \citet{o2017gulf}. Thirdly, a measure of the North Atlantic ``cold SST bias'', computed as the mean SST over the box 40-50N, 45-30W. This box roughly encloses the large cold bias visible in Figure \ref{fig:sst_biases}(f). This bias has been argued to influence blocking/jet variability by \citet{scaife2011improved} and \citet{Keeley2012}: it is sometimes colloquially referred to as the ``blue blob of death''. Our choices are motivated by the desire to test metrics that are easily interpretable and corroborated by prior literature. Thus we do not test metrics such as the basin-wide root-mean square error (RMSE) of North Atlantic SSTs, as done in \citet{davini2020cmip3}, as these cannot be clearly interpreted: the biases contributing to the error could in principle differ widely from model to model. Note that both SST gradient metrics are computed by taking zonal averages of the SSTs over the respective region and regressing the zonally averaged SSTs against latitude: the regression coefficient is defined to be the SST gradient.

The resulting scatter plots are shown in Figure \ref{fig:sst_scatter}. In all three cases, correlations between the SST metric and European blocking/NorthDay frequencies are found to be low. For European blocking, the correlations, in order, are $-0.22. -0.16$ and $-0.12$, while for NorthDays they are $-0.03$, $-0.02$ and $0.07$. The only significant correlation is the one between the Gulf Stream extension SST gradient and European blocking. This suggests SST gradients in this region are robustly associated with blocking frequencies, consistent with the hypothesis of \citet{athanasiadis2022mitigating}. However, the SST gradient correlation of $-0.22$ is considerably lower in magnitude than the one we report between Gulf Stream precipitation variability and blocking of $0.67$, and the SST gradient correlation of $-0.03$ is negligible in the case of NorthDays, unlike the correlation of $0.57$ reported using precipitation variability. It is worth emphasising that the SST gradient correlations are notably higher if one restricts to only HighResMIP models, in line with the consistency of the relationship reported by \citet{athanasiadis2022mitigating}: see Supplemental Figure S4. There is some evidence to suggest that HighResMIP may to some extent be systematically different from the wider multimodel ensemble, at least in the case of European blocking, but it also seems likely that the stronger link between SST gradients and blocking frequencies in HighResMIP is inflated by sampling variability, and that the true link is lower, as in our multimodel ensemble. See Supplemental Text S1 for further discussion.

Thus, based on the three metrics we considered, it seems that North Atlantic SST biases are at best weakly associated with biases in European blocking and northern jet excursions. This is not to say SST biases are unimportant. For any \emph{specific} model, its biggest SST bias could contribute more strongly, as suggested by the correlation between RMSE and blocking reported by \citet{davini2020cmip3}. Our analysis suggests, however, that there may not be any systematic SST bias across all models that explains biases in blocking/jet variability. In our view, a more likely explanation is that reductions in SST biases in coupled models may so far have simply been too modest to have a big effect. Figure \ref{fig:sst_scatter}(e) suggests that the vast majority of models still have a North Atlantic cold-spot bias of around 2 or more degrees, supporting this possibility.

The possibility that reductions in North Atlantic SST biases have so far been very modest in coupled models is arguably not surprising. The key source of ocean-driven biases in the North Atlantic arise from a failure to simulate the path of the Gulf Stream correctly. While \citet{michel2023increased} argue there are some improvements to the simulated trajectory with increased ocean resolution, \citet{tsartsali2022impact} conclude that, of all the HighResMIP models they consider, only the one model using a $1/12$-th degree ocean is able to simulate a more realistic Gulf Stream separation. The evidence that biases in CMIP models have been significantly reduced due to a more realistic Gulf Stream therefore seems unclear given that virtually no models to date have run at such a high ocean resolution. The low/negligible correlations we find here (Figure \ref{fig:sst_scatter}) corroborate this ambiguity.

\subsection{Coupled vs AMIP models, and the relative roles of atmospheric and oceanic resolution}
\label{sec:coupled_vs_amip}

Both \citet{athanasiadis2022mitigating} and \citet{michel2023increased} argue that the improvements seen in blocking/jet variability in coupled models with increased model resolution is primarily a result of increased ocean resolution. An important quantitative argument rests in both cases on the observation that the bias reduction associated with increased resolution appears to be smaller in AMIP models than in coupled models, which has been interpreted as evidence for the dominant role of the ocean resolution. On the other hand, it is noteworthy that AMIP models by and large have similar biases in blocking to coupled models \citep{schiemann2020northern}, meaning that reducing the SST biases to zero does not actually eliminate blocking biases. This strongly suggests that high oceanic resolution alone is not sufficient, and increased atmospheric resolution must also be playing some role in alleviating biases. Indeed, \citet{schiemann2020northern} suggest that the similar biases in AMIP and coupled models may indicate the importance of air-sea coupling which requires high oceanic \emph{and} atmospheric resolution \citep{moreton2021air}.

But if atmospheric resolution plays any role whatsoever at alleviating biases, then it follows that the improvement with resolution must be smaller in AMIP models than in coupled models. This is because in coupled models the resolution increase happens almost always in both the ocean and the atmosphere, and so the increased resolution in coupled models will be associated with an alleviation of both the atmospheric \emph{and} oceanic resolution-related biases, while in AMIP models the only biases alleviated are ones related to the atmospheric resolution. This difference is likely exacerbated by the suppressed precipitation variability in AMIP models that we reported in Section \ref{sec:results}\ref{sec:thermal_damping}, especially given the potentially non-linear relationship between atmospheric resolution and precipitation variability. In other words, increasing the atmospheric resolution in an AMIP model may have a much more modest impact on precipitation variability compared to the same increase in a coupled model. Even ignoring any potential non-linearity, AMIP models at 25km atmospheric resolution still have precipitation variability well below that of 100km resolution coupled models (Figure \ref{fig:amip_vs_ocean}), meaning high-resolution AMIP models still vastly underestimate the real-world variability. Consequently, the bias reduction in European blocking and northern jet excursions may be more modest in AMIP models.

The fact that AMIP models have lower precipitation variability but similar blocking frequencies seems at first glance to contradict the importance of Gulf Stream precipitation. Instead we argue that this may actually help explain the puzzle about why AMIP and coupled models have similar biases. In AMIP models, the SST biases have been eliminated, but the lack of coupling kills the precipitation variability. These factors may cancel each other out and result in no overall change in blocking biases. Another important factor may be the broader role played by air-sea coupling in blocking formation and maintenance, as laid out in \citet{mathews2024oceanic}, though it is not clear to us if the lack of coupling in AMIP models would enhance or suppress the mechanisms they describe. What is in any case clear is that a direct comparison between AMIP and coupled models is not straightforward.

We thus argue that it is hard to conclude that oceanic resolution has played a more important role than atmospheric resolution in models to date. It does not even seem easy to rule out the possibility that the effect of increased ocean resolution has so far been minimal. After all, increased atmospheric resolution by itself will enhance precipitation variability \citep{scher2017resolution, schemm2023toward}, and improvements to blocking/jet variability would themselves be expected to reduce SST biases, via air-sea coupling. Previously reported correlations between blocking/jet variability and SST biases could therefore be mediated by Gulf Stream precipitation variability which in turn may be set primarily by the atmospheric resolution. The results of the previous subsection could be interpreted as corroborating this possibility. More detailed analysis here goes beyond the scope of the present paper, but is of clear interest for future work.

\subsection{Shortcomings of our work}
\label{sec:shortcomings}

A key shortcoming of our analysis is the lack of process-based analysis. In particular, we have not attempted to further our detailed understanding of the relevant physical processes involved in translating Gulf Stream precipitation to European blocking and northern jets, and how these processes are simulated in models. One prominent way this shortcoming affects our results is the uncertainty introduced by the effect of different convection parameterisations in models. Figure \ref{fig:scatter_resolution}(a) shows that, for a fixed horizontal resolution, different subsets of the ensemble sometimes appear to cluster around very different values of GS Precip. While we cannot rule out a potential influence of decadal modes of variability, which will vary across the models, a pragmatic explanation is that these differences are largely due to differences in the convection parameterisation. Such a sensitivity was also reported in \citet{strandberg2021importance}. If Gulf Stream precipitation variability really is a primary source of intermodel spread in blocking/northern jets, then the similar `clustering' behaviour seen in Figure \ref{fig:scatter_resolution}(b) and (c) at different fixed resolutions may also be partially due to differences in the convection parameterisations. Accounting for these differences would probably require the kind of careful process-based analysis that our paper omits. However, our work here offers a complementary view to the increasing body of such detailed studies, by showing that even quite crude measures suffice to establish robust correlations in multimodel ensembles. It is therefore possible that a more carefully chosen metric of the relevant diabatic processes would result in even higher correlations than those seen in Figure \ref{fig:scatter_precip}.

Another clear weakness of our analysis is that our choice to pool all models together might obscure genuine differences across, e.g., successive generations of climate models. There is, for example, no reason \emph{a priori} for the biases in CMIP6 and CMIP5 to share the same underlying cause. Our results show that Gulf Stream precipitation variability could be such a common cause, but the correlations we computed would, viewed in isolation, not be sufficient evidence to draw a strong conclusion. However, when viewed in combination with the several targeted experiments and case studies assessing the role of diabatic processes in the Gulf Stream on blocking, it becomes more plausible that our analysis is highlighting a genuine relationship.

Finally, when assessing the role of horizontal resolution, we used the nominal grid resolution as our index, which tends to be a systematic overestimate of the effective resolution \citep{klaver2020effective}. This could potentially bias our analysis. It would be valuable if modelling centres offered an explicit estimate of the effective resolution of all their CMIP-related simulations as a matter of course.

\section{Summary and Conclusions}
\label{sec:conclusions}

We present a brief synopsis of our results:
\begin{enumerate}
    \item A simple measure of climatological Gulf Stream precipitation variability is shown to be strongly associated with both climatological European blocking frequency (correlation of 0.67) and northern jet days (correlation of 0.57) in a large coupled multimodel ensemble (Figure \ref{fig:scatter_precip}). In the case of European blocking, as much as 45\% of the intermodel spread is associated with variations in precipitation variability. Models experience particularly severe biases in precipitation variability.
    \item Gulf Stream precipitation variability goes up with atmospheric horizontal resolution: Figure \ref{fig:scatter_resolution}, possibly in a super-linear manner. Statistical tests for mediation corroborate a hypothesis that the reduced biases in blocking and northern jet days seen with increased model resolution are mediated by enhanced precipitation variability. In the model ensemble, a day with heavier rainfall in the Gulf Stream region is followed by a stronger downstream wave-train (Figure \ref{fig:composites}).
    \item AMIP models exhibit systematically reduced precipitation variability compared to coupled models, due to the ``excess thermal damping'' mechanism described by \citet{barsugli1998basic} (Figure \ref{fig:amip_vs_ocean}). This complicates comparisons between AMIP and coupled models.
    \item Important SST biases, such as Gulf Stream SST gradients or the North Atlantic cold SST bias, are found to share only weak or negligible correlations with European blocking frequencies and northern jet days (Figure \ref{fig:sst_scatter}).
\end{enumerate}
We interpret our results as corroborating the hypothesis, put forth in \citet{schemm2023toward}, that biases in diabatic processes in the Gulf Stream region are a key source of biases in the North Atlantic eddy-driven jet, and furthermore that increasing the atmospheric horizontal resolution of models reduces jet biases by better simulating these diabatic processes. More broadly, our work shows that diabatic processes in the Gulf Stream region are not just important for understanding variability of particular events \citep{wenta2024linking} or in particular models \citep{mathews2024gulfstream}, but also for the question of understanding intermodel spread and the impact of model resolution. We have also highlighted the value of using metrics measuring the variability of precipitation, as opposed to the mean, in order to better capture the intermittent nature of storm track activity. Our results generally seem to hint that the inability of models to simulate more extreme rainfall may be an important source of biases in blocking and the jet.

While previous studies have tended to focus on the role of SST biases and oceanic horizontal resolution, studies such as \citet{scher2017resolution} and \citet{schemm2023toward} show that atmospheric resolution alone can improve both precipitation variability and the jet. Given that (a) we find a weak or negligible link between SST biases and jet variability in our ensemble, and (b) previous arguments based on a comparison of AMIP and coupled models are problematic in light of the systematically suppressed precipitation variability experienced by AMIP models, we speculate that improvements to metrics like European blocking frequency with model resolution, reported in \citet{davini2020cmip3} and reproduced here, are primarily due to increased atmospheric resolution. We do expect that reducing the SST biases also helps, but our analysis suggests that coupled models have to date only experienced a modest reduction of SST biases, likely owing to the apparent need for an oceanic resolution of $1/12$th of a degree or higher in order to simulate a genuinely realistic Gulf Stream separation \citep{tsartsali2022impact}.

It must be emphasised at this point that the Gulf Stream \emph{per se} is conspicuously absent in our analysis. That is, we have looked at the precipitation variability in the Gulf Stream region, but we have not linked this variability to, e.g., oceanic heat transport or the SST gradients created by the Gulf Stream. Consideration of lagged influences of precipitation with such quantities could help clarify the role played by the Gulf Stream itself, beyond just being a source of warm water for the atmosphere to feed on. The work of \citet{mathews2024oceanic} strongly suggests that one should view the system as a fundamentally coupled one, and a natural avenue for future work is to look at how this coupling is simulated in models.

We end with some thoughts about what might happen when we reach the kilometre scale models championed by \citet{slingo2022ambitious} and others. This will at least partially depend on whether IMERG or TRMM is a better estimate of reality. If TRMM is the better estimate, then a considerable step change increase in precipitation variability may well take place. One interesting possibility here is that resolving mesoscale eddies in the ocean may profoundly impact air-sea fluxes in models, as hinted at by recent work \citep{yang2024observations}. This could end up playing an important role in any such step changes. Less clear is what the impact will be of models simulating a realistic Gulf Stream (including the ``northwest corner'') \emph{en masse}, since this could expose the effect of compensating biases or tuning efforts made to suppress the effect of North Atlantic SST biases. Here too though earlier work suggests that ultra-high resolution models may behave quite differently \citep{famooss2022atmospheric}. Coupled model experiments aimed at understanding the impact of resolving mesoscale eddies on the climate system is the goal of the ``EERIE'' Project (doi:10.3030/101081383). One may therefore hope that some of these questions will begin to be resolved over the coming years.

%

%

\clearpage
\acknowledgments

This publication is part of the EERIE project (Grant Agreement No 101081383) funded by the European Union. Views and opinions expressed are however those of the author(s) only and do not necessarily reflect those of the European Union or the European Climate Infrastructure and Environment Executive Agency (CINEA). Neither the European Union nor the granting authority can be held responsible for them. University of Oxford's contribution to EERIE is funded by UK Research and Innovation (UKRI) under the UK government’s Horizon Europe funding guarantee, grant number 10049639. All three authors therefore acknowledge funding from this grant. HMC was further funded by Natural Environment Research Council grant number NE/P018238/1.

KS thanks Fenwick Cooper for several helpful conversations about precipitation in models versus reality, as well as Matthew Patterson for helpful feedback on an initial draft.

We acknowledge the World Climate Research Programme, which, through its Working Group on Coupled Modelling, coordinated and promoted CMIP6. We thank the climate modeling groups for producing and making available their model output, the Earth System Grid Federation (ESGF) for archiving the data and providing access, and the multiple funding agencies who support CMIP6 and ESGF.

For the purpose of Open Access, the author has applied a CC BY public copyright licence to any Author Accepted Manuscript (AAM) version arising from this submission."

%
%
\datastatement

ERA5 data is publicly available via the Copernicus Data Store \citep{era5_dataset}.

IMERG version 07 data is available courtesy of NASA. The IMERG Version 07 data were provided by the NASA/Goddard Space Flight Center's precipitation processing center, which develop and compute IMERG as a contribution to GPM, and archived at the NASA GES DISC \citep{imergv07_dataset}.

CMIP6 data \citep{eyring2016overview} can be freely downloaded from the ESGF at \url{https://esgf-node.llnl.gov/projects/cmip6/}. CMIP5 and HighResMIP data can similarly be downloaded from the ESGF.

%

\appendix

We give an informal and brief description of the statistical model of \citet{barsugli1998basic} (hereafter BB98) used to represent a coupled atmosphere-ocean system and an AMIP-style counterpart. We do this in order to keep the present paper self-contained, and to draw renewed attention to the importance of BB98 for studies comparing coupled and AMIP models. Readers should refer to BB98 for more details, including how the parameter values were chosen.

Let $T_a$ denote anomalous near-surface air temperatures, and $T_o$ anomalous sea-surface temperatures, both thought of as being averaged over a generic North Atlantic domain. Then BB98 model the evolution of $T_a$ and $T_o$ as a system of coupled stochastic differential equations:
\begin{eqnarray}
        &\frac{d}{dt}T_a& = \,A\cdot T_a\,+\, B\cdot T_o \,+\, \epsilon,\\
        &\frac{d}{dt}T_o& = \,C\cdot T_a\,+\, D\cdot T_o.
\end{eqnarray}
Here $A, B, C$ and $D$ are constant coefficients and $\epsilon$ is a stochastic noise term assumed to be normally distributed with mean zero. Such models have been used extensively in the literature to capture atmosphere-ocean coupling (see e.g., \citet{PeSa95, Penland1993, Alexander2008, Hawkins2009, Newman2009} for some examples). The coefficients $A$ and $D$, which are necessarily negative in a stable system, represent damping terms which relax $T_a$ and $T_o$ back to their mean (i.e., zero), while the coefficients $B$ and $C$ account for coupling.

Using the notation from BB98, we have $A= -(\lambda_{sa}c + \lambda_a)/\gamma_a, B=b\lambda_{sa}/\gamma_a, C=c\lambda_{so}/\gamma_o, D=-(\lambda_o+\lambda_{so})/\gamma_o$. The unknown parameters here are given in Table 1 of BB98 as follows: $\gamma_a = 1\cdot 10^{-7}$, $\gamma_o = 2\cdot 10^{-8}$, $\lambda_{sa} = 23.9$, $\lambda_{so} = 23.4$, $\lambda_a = 2.8$, $\lambda_o = 1.9$, $c=1$ and $b=0.5$. After rescaling the entire system by $10^6$, this gives $A=-2.67, B=1.195, C=0.117$ and $D=-0.126$. The standard deviation of the noise in BB98 is given as $\lambda_{sa}/\gamma_{a}$, which after rescaling by $10^6$ is $2.39$.

\begin{figure}[ht!]
\centering
  \noindent\includegraphics[width=25pc,angle=0]{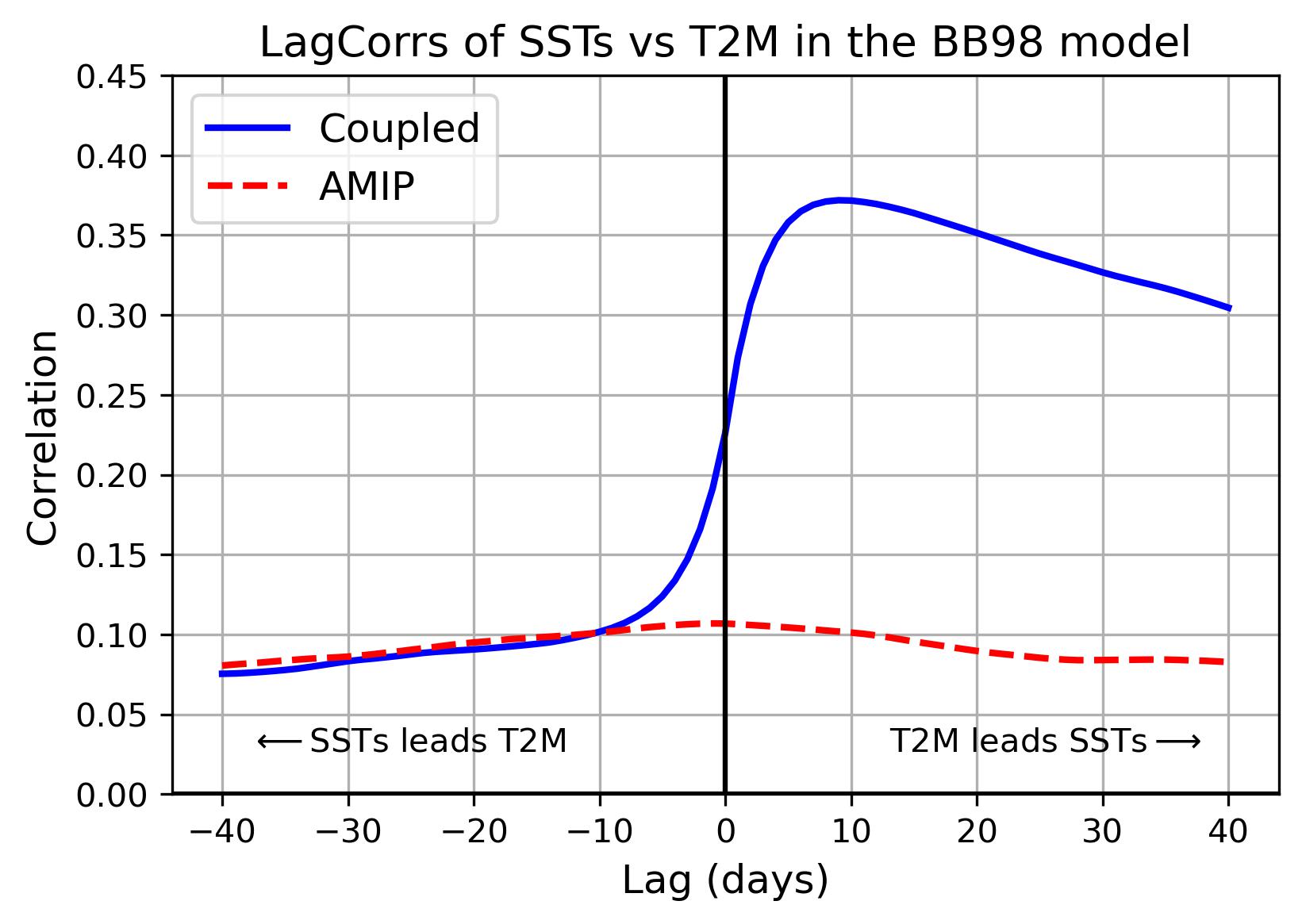}\\
  \caption{Lead-lag correlations of daily SST vs daily T2M for based on the BB98 model; the coupled version is plotted in solid blue and the AMIP-style version using stipled red. Negative lags correspond to SSTs leading T2M, and positive lags to T2M leading SSTs.}\label{fig:bb98}
\end{figure}

One may now readily simulate a long timeseries of the fully coupled system by integrating (A1) and (A2), yielding in particular a timeseries of simulated SSTs, $T_{o}^{CPL}$. To generate an AMIP-style counterpart, following BB98 we simply integrate equation (A1) where at every timestep $t$ we put $T_o(t) = T_{o}^{CPL}(t)$. We carried this out, integrating the system for 50,000 timesteps, and then computed lag correlations of SSTs ($T_o$) and T2M ($T_a$) for the coupled and AMIP-style systems. The result is shown in Figure \ref{fig:bb98}, reproducing one of the key results of BB98. Comparison with Figure \ref{fig:amip_vs_ocean}(b) confirms the conclusions drawn already in BB98, that lack of coupling in a model has a profound impact on air-sea interactions, not just when T2M leads SSTs (which is a priori obvious) but also when SSTs lead T2M (see the discrepancy between the blue and red lines for lags between -10 and 0).

BB98 furthermore show that the lack of coupling leads to a systematic reduction in daily T2M variability. B98 explain this result, as well as the change in lag-correlations, as being due to the excessive thermal damping experienced by AMIP models. In the coupled system, if $T_a$ and $T_o$ are positive (negative) on a given timestep, the coupling terms ($BT_o$ and $CT_a$) will act to try to persist the positive (negative) anomalies on the next timestep as well, opposing the effect of the damping terms (which act to remove the anomaly), thereby increasing the likelihood of very persistent anomalies occurring. The coupling terms thus serve to enhance the low-frequency variability of both $T_a$ and $T_o$; when this coupling is lost, such as in AMIP-style simulations, this additional variability is lost.

A heuristic ``physical'' explanation for this mathematical effect comes from consideration of heatfluxes. The thermal component of the anomalous heatflux is proportional to $T_a - T_o$, up to some sign convention. If one assumes that $T_o$ is constant and zero (as in a simulation using fixed climatological SSTs), this implies that the sign of the anomalous thermal heatflux is determined purely by the sign of $T_a$, and will thus always act to oppose the anomaly. If $T_a>0$, the anomalous heatflux will point down (i.e., the atmosphere will, in an anomalous sense, donate extra heat to the ocean), while if $T_a<0$, the anomalous heatflux will point up (the atmosphere will extract extra heat from the ocean). In AMIP-style simulations, the SSTs are not fixed, but because the SSTs are varying independently of T2M, the same heuristic essentially applies \emph{on average}. This is the sense in which BB98 refer to AMIP-models as experiencing excessive thermal damping.





%



\bibliographystyle{ametsocV6}
\bibliography{references}

\end{document}